\documentclass[preprint,12pt]{elsarticle}

\usepackage{epsfig}
\usepackage{amssymb}
\usepackage{amsmath}
\usepackage{multirow}
\usepackage{hyperref} 

\def\pt{\ensuremath{p_{\mathrm{T}}}} 
 
\def\et{\ensuremath{E_{\mathrm{T}}}} 
 
\def\abseta{\ensuremath{|\eta|}}

\journal{Physics Letters B}

\begin{document}

\begin{frontmatter}

\title{\vspace{-0.5cm} \flushright{\normalsize{CERN-PH-EP-2011-108}} \\ \center{Measurements of the electron and muon inclusive cross-sections in proton-proton collisions at $\sqrt{s}$~=~7~TeV with the ATLAS detector}}

\author{The ATLAS collaboration}

\begin{abstract}
This letter presents measurements of the differential cross-sections for
inclusive electron and muon production in proton-proton collisions at
a centre-of-mass energy of $\sqrt{s} = $~7~TeV, using data collected  by the ATLAS detector at the LHC.
The muon cross-section is measured as a function of~\pt~in the range $ 4 < \pt < 100$~GeV
and within pseudorapidity $|\eta| < 2.5$. In addition the electron and muon cross-sections are measured
in the range $ 7 < \pt < 26$~GeV and within $|\eta| <2.0$, excluding $1.37<|\eta|<1.52$. Integrated
luminosities of 1.3 pb$^{-1}$ and 1.4 pb$^{-1}$ are used for the electron and muon measurements,
respectively.
After subtraction of the $W$/$Z$/$\gamma^{\ast}$ contribution, the
differential cross-sections  are found to be in good
agreement with theoretical predictions for heavy-flavour production
obtained from Fixed  Order NLO calculations with NLL high-\pt~resummation, and to be sensitive to the effects of NLL resummation. 
\end{abstract}


\end{frontmatter}

\section{Introduction}

An understanding of electron and muon 
production in proton-proton ($pp$) collisions is a prerequisite for measurements and searches  
including these particles in the final state. 
Moreover, 
the inclusive production of these particles can be used to constrain
theoretical predictions for heavy-flavour production, for which large uncertainties exist.
At low transverse momentum (\pt) the inclusive electron and muon spectra are dominated by decays of charm and beauty hadrons. 
The contribution from $W$ and $Z/\gamma^{\ast}$ production,
which dominates in the higher \pt~region, is well-understood~\cite{firstWZ} and may be 
subtracted in order to obtain the heavy-flavour cross-section.

In measurements of $b$-quark production in ${p}\bar{{p}}$  collisions,  an excess over the
theoretical expectation was observed in earlier experiments~\cite{UA1,D0,CDF,cdfele}.
This discrepancy was later resolved by improved experimental measurements~\cite{CDFnew} and the use of Next to Leading Order 
(NLO) with Next to Leading Log (NLL) resummation
theory applied to LEP data to extract the $b$-quark fragmentation
function~\cite{Cacciari_1,Mangano}.
The Tevatron data were, however, not sensitive to the \pt~region where  the deviation between the NLO and the NLO + NLL 
perturbative QCD (pQCD) calculations becomes apparent.
At the LHC, NLL resummation can be probed directly in the pQCD
prediction for heavy-flavour production in hadron collisions for the
first time.

In the analyses reported in this letter the \pt~spectra of inclusive electrons and muons
are measured  using an integrated luminosity of~$1.3$~pb$^{-1}$  and
$1.4$~pb$^{-1}$, respectively. 
A kinematic acceptance of  $ 7 < \pt < 26$~GeV 
and pseudorapidity{\footnote {ATLAS uses a right-handed 
coordinate system with its origin at the nominal interaction point (IP) in the centre of the
detector and the z-axis coinciding with the axis of the beam pipe. 
The pseudorapidity is defined in terms of the polar angle $\theta$ 
as $\eta = - \rm {ln}~\rm{tan}(\theta/2)$.}}
$|\eta|<2.0$ excluding $1.37<|\eta|<1.52$ is considered for electrons, and 
$4 < \pt < 100$~GeV and $|\eta|<2.5$ for muons.

 This letter is organised as follows. The experimental and theoretical methodology
is  outlined
in Section 2. A short description of electron and muon reconstruction in the ATLAS detector
is provided in Section 3, with the recorded and simulated data samples used in the analyses being discussed in
Section 4. Sections 5 and 6 describe the cross-section measurements in the
electron and muon channels respectively.
For the muon analysis, the inclusive cross-section is compared to the most recent theoretical predictions in Section 6.6.
Finally in Section 7, the 
electron and muon cross-sections from heavy-flavour hadron
production are determined by subtracting
the $W$/$Z$/$\gamma^{\ast}$ contributions.
These results are compared to the predictions of NLO+NLL and NLO calculations using the program FONLL~\cite{FONLL-HQ0,FONLL-HQ1}.
Comparisons are also made to the NLO predictions from the {\tt POWHEG}~\cite{Powheg1,Powheg3} program
and the Leading Order (LO) expectations from {\tt PYTHIA}~\cite{Pythia}.

\section{Cross-section measurement and theoretical predictions}

The measured differential cross-section within the kinematic acceptance of the charged lepton is defined by
\begin{equation}
\frac{\Delta\sigma_i}{\Delta p_{{\rm {T}}_i}} =  \frac{N_{\mathrm{sig}_{\,i}}}{\Gamma_{{\rm{bin}}_{\,i}} \cdot \int{\cal L} dt} 
\cdot 
\frac{C_{\mathrm{migration}_i}}{\epsilon_{\mathrm{(reco+PID)}_i} \cdot \epsilon_{\mathrm{trigger}_i}},
\label{eq:cross_section}
\end{equation}
where 
$N_{\mathrm{sig}_{\,i}}$ is the number of signal electrons or muons with reconstructed
$p_\mathrm{T}$ in bin $i$ of width ${\Gamma_{{\rm{bin}}_{\,i}}}$,
$\int{\cal L}dt$ is the integrated luminosity, 
$\epsilon_{\mathrm{trigger}_i}$ is the trigger efficiency and
$\epsilon_{\mathrm{(reco+PID)}_i}$ is the combined reconstruction and identification efficiency.
$C_{\mathrm{migration}_i}$ is the bin migration correction factor, 
defined as the ratio of the number of charged leptons in bin $i$ of true
\pt~and the number in the same bin of reconstructed \pt~(transverse energy, \et, in the electron case).
The methods used to extract $N_{\mathrm{sig}_{\,i}}$  from the total number of electron or muon candidates observed in each \pt~bin are explained in Sections 5.3 and 6.4.
From the extracted signals, we subtract the contribution from $W$/$Z$/$\gamma^{\ast}$ production in order to obtain
a cross-section corresponding to the  decays of heavy-flavour hadrons produced in the $pp$ collisions to electrons or muons.
In the electron analysis, the $W$/$Z$/$\gamma^{\ast}$ accepted cross-section,
$\sigma^{\mathrm{W/Z/\gamma^{\ast}}}_{\mathrm{accepted}_i}$, is subtracted 
before applying the efficiency and migration correction factor, 
$\epsilon_{\mathrm{{(reco+PID)}}_i}/C_{\mathrm{migration}_i}$,
which is specific to heavy-flavour electrons due to the dependence of the identification efficiency on isolation. 
In the muon analysis, the same correction factor applies for 
muons originating from both heavy-flavour and $W/Z/\gamma^{\ast}$ decays, allowing the subtraction to be performed
at the cross-section level.

The spectrum of charged leptons from heavy-flavour 
decays is calculated in a theoretical framework, FONLL,
permitting direct comparison with the data.
FONLL is based on three main components: the heavy quark production cross-section calculated in pQCD by matching 
the Fixed Order NLO terms with NLL 
high-\pt~resummation, the non-perturbative heavy-flavour fragmentation functions determined from  
$e^+e^-$ collisions and extracted in the same framework, 
and the decays of the heavy hadrons
 to leptons using decay tables and form factors from $B$-factories.  
The theoretical uncertainties associated with the FONLL prediction will be discussed in Section 7 when the comparisons
to the measured cross-sections are made.

\section{Electron and muon reconstruction in the ATLAS detector}

The ATLAS detector consists of three main components: an  Inner Detector (ID) tracking system immersed in
a 2~T magnetic field, surrounded by electromagnetic (EM) and hadronic calorimeters and an
outer muon spectrometer (MS). A full description can be found in~\cite{DetectorPaper}. 
The ID provides precise track reconstruction within
 $|\eta| < 2.5$, employing pixel detectors close to the beam-pipe, 
silicon microstrip detectors~(SCT) at intermediate radii and a Transition Radiation Tracker~(TRT) at
outer radii.
Within $|\eta| < 2.0$ the TRT provides substantial discriminating power between
electrons and pions over a wide energy range.
The inner-most pixel layer (the B-layer) is located at a radius of~50~mm and provides precision vertexing and
significant rejection of tracks produced by photon conversions.

Within~$\abseta < 2.5$, EM calorimetry is provided by the 
barrel and end-cap lead/Liquid-Argon (LAr) EM sampling calorimeters,
and hadronic calorimetry by the three-part steel/scintillating tile barrel calorimeter plus the 
two copper/LAr end-caps. 
The EM calorimeter is segmented in the longitudinal and transverse directions, with fine granularity along the $\eta$ 
direction in the first (strip) layer.
The identification of electron candidates is seeded by a  preliminary set of clusters in the EM
calorimeter using a sliding window algorithm, with those clusters having a match
to a suitable ID track being reconstructed~\cite{CONFNote2009}.
In the transition region between the barrel and end-cap calorimeters at $1.37 < |\eta| < 1.52$
the electron identification and energy resolution is degraded by the large amount 
of material in front of the first active layers, prompting the exclusion of this region from the electron analysis.

The MS comprises separate trigger and high-precision tracking chambers which measure the deflection of muons in a magnetic field generated
by three super-conducting air-core toroids. The precision chamber system covers the region~$\abseta < 2.7$ with three layers of Monitored
Drift Tube (MDT) chambers. 
In the forward region, $2.0 < |\eta| < 2.7$,  higher granularity Cathode Strip Chambers (CSCs) replace the first station of MDTs.
The trigger chambers provide coverage within~$|\eta| < 1.05$ using Resistive Plate Chambers
(RPCs) and for $1.05 < |\eta| < 2.4$ using Thin Gap Chambers (TGCs).
The MDT chambers measure the coordinate in the
bending plane, while the RPCs and TGCs
measure the coordinate in the non-bending plane ($\phi$) and provide a further hit in the bending plane. 

Reconstruction of muon candidates begins with the reconstruction of
track segments in the MS.  Segment candidates formed from hits in the
precision chambers are
required to point loosely to the centre of ATLAS. A minimum of  two track 
segments and one hit in each coordinate of the RPCs in the barrel and the TGCs
in the end-caps are required to build an MS track.
For $|\eta| < 2.5$ the track parameters are then back-extrapolated to
the IP and  matched to all tracks in the ID
having hits in at least two ID sub-detectors.
 The ID track that best matches the MS track is retained, and the track 
parameters are computed by the statistical combination of back-extrapolated MS parameters and ID track parameters, 
the resulting track being referred to as a combined muon in the following.

\section{Data and simulated samples used}

The analysis is based on a data sample collected at $\sqrt{s} = 7$~TeV  during April-August 2010.
Requirements were made on the detector conditions (notably the ID plus either the EM calorimeter or the MS)
and data quality, yielding total integrated luminosities of~$1.28\pm 0.04$~pb$^{-1}$  and $1.42\pm 0.05$~pb$^{-1}$ for
the electron and muon analyses, respectively, the integrated luminosity being measured with an
uncertainty of 3.4\%~\cite{LumiSyst}.

For the electron analysis events were selected using the hardware-based first-level (L1) calorimeter trigger, which identifies EM clusters 
within $|\eta|< 2.5$ above a given energy threshold.
The data were recorded under four different trigger conditions, with
a progressively higher minimum 
cluster transverse energy requirement applied as the instantaneous luminosity of the LHC increased.
The bulk of the integrated luminosity (76\%) was obtained with the L1 calorimeter trigger configured with an 
energy threshold 
of approximately 15~GeV, with the remaining 14\%, 9\% and 1\% recorded with 11, 6 and 3~GeV thresholds, respectively. 
The integrated luminosity available for the electron analysis is limited to these early data, 
since the Higher Level Trigger algorithms used in later periods of higher instantaneous luminosity are designed to be 
efficient only for isolated electrons.

In the muon channel, events were selected by one of two L1 muon triggers.
The first 3.5\% of the data were recorded under the loosest requirement of 
at least three trigger hits in time coincidence with the collision
(referred to as the lower threshold trigger), while the remaining data were obtained with the further requirement that the
hit pattern be compatible with a track with \pt~$> 10$ GeV.
In the subsequent analysis it is required for muons with \pt~less than 16 GeV to be triggered by the 
lower threshold trigger, while the 10~GeV
trigger is required for muons with \pt~in the range 16-100~GeV.

Simulated data samples have been generated in order to estimate backgrounds 
and correct for the trigger and reconstruction efficiencies and 
the resolution of the detector.
{\tt PYTHIA 6.421} was used to simulate samples of electrons and muons
from heavy-flavour and $W$/$Z$/$\gamma^{\ast}$  decays.
{\tt PYTHIA} was also used to simulate all sources of background electrons and muons.
Further samples of electrons from heavy-flavour decays were also generated with {\tt POWHEG-hvq v1.0 patch 4},
interfaced to either {\tt PYTHIA} or {\tt HERWIG v6.510}~\cite{herwig}.
In conjunction with {\tt HERWIG}, {\tt JIMMY v4.31}~\cite{jimmy} was used to model the
underlying event.
 The {\tt POWHEG} samples 
use {\tt PHOTOS v2.15}~\cite{photos} to model final state QED radiation. 
The PDF set used was {\tt MRST LO*}~\cite{mrst2007lomod} for the {\tt PYTHIA} samples and  {\tt CTEQ6.6}~\cite{cteq6l1}
for the {\tt POWHEG} samples.
All signal and background samples were 
generated at $\sqrt {s} = 7$~TeV using the ATLAS MC09 tune~\cite{mc09Tune}, 
and passed through the {\tt GEANT4}~\cite{geant4} simulation
of the ATLAS detector.

\section{Electron analysis}

\subsection{Electron candidate selection}

Events from $pp$ collisions are selected by requiring a collision vertex with more than two associated tracks.
From these events, reconstructed electron candidates are required to
pass a minimum cluster \et~cut
between 7 and 18~GeV depending on the
trigger condition, to lie within the pseudorapidity coverage of the~TRT, $|\eta| < 2.0$, and to be 
outside the transition region between the barrel and end-cap calorimeters, $1.37 < |\eta| < 1.52$.
Candidate clusters with their energy-weighted centre close to 
 problematic regions in the EM~calorimeter are rejected,
as are those with tracks passing through dead B-layer modules:
the corresponding loss of acceptance varied by run period
but amounted to no more than 7\% and 3\%, respectively.

{\em Preselected} candidates must be associated to tracks containing at least ten TRT and
four silicon hits and are required to pass a minimum
requirement on the fraction of the raw energy deposited in the strip layer of the EM calorimeter.
Candidate electrons are then selected from those passing the preselection
by imposing further {\em identification} criteria~\cite{CONFNote2009} designed to suppress electron-like (fake) 
signatures from hadrons. 
These identification criteria comprise \et~ and $|\eta|$ dependent cuts on the energy deposits in the strip and
 middle layers of the EM calorimeter as well as on the track quality and track-cluster matching.

\begin{figure}[h!]
\begin{center}
\begin{tabular}{cc}
\includegraphics[width=0.48\textwidth]{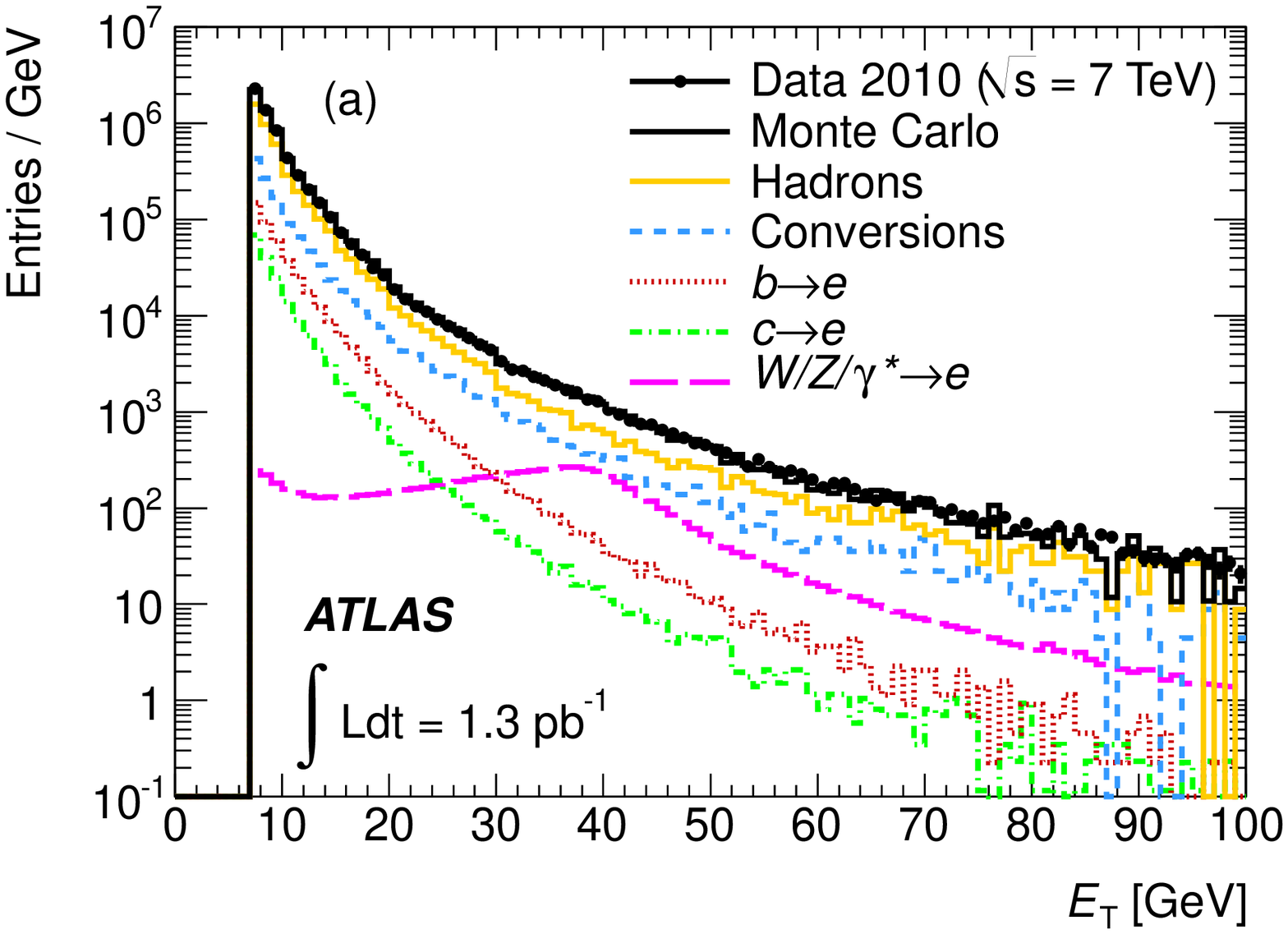} &
\includegraphics[width=0.48\textwidth]{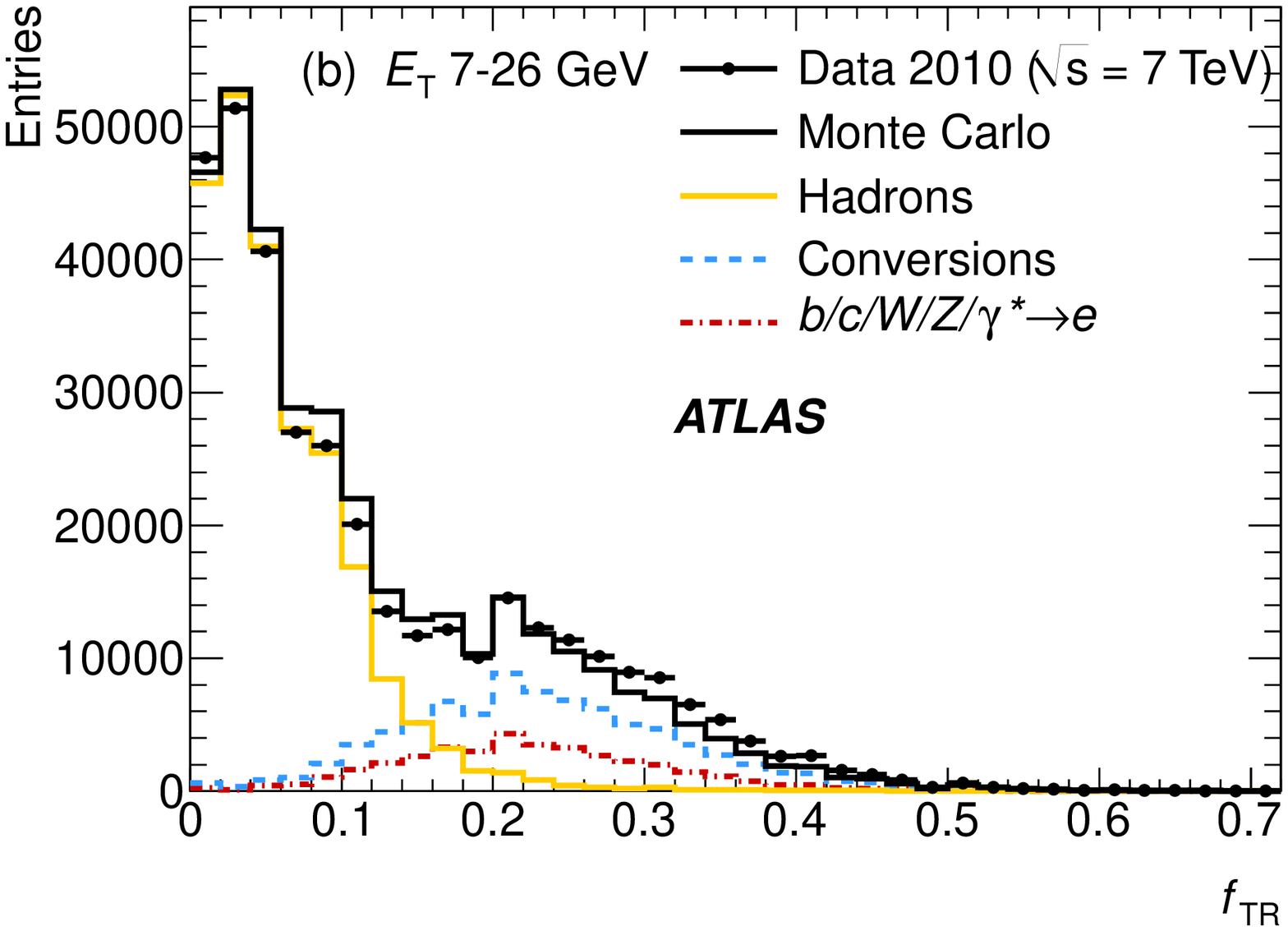} \\
\includegraphics[width=0.48\textwidth]{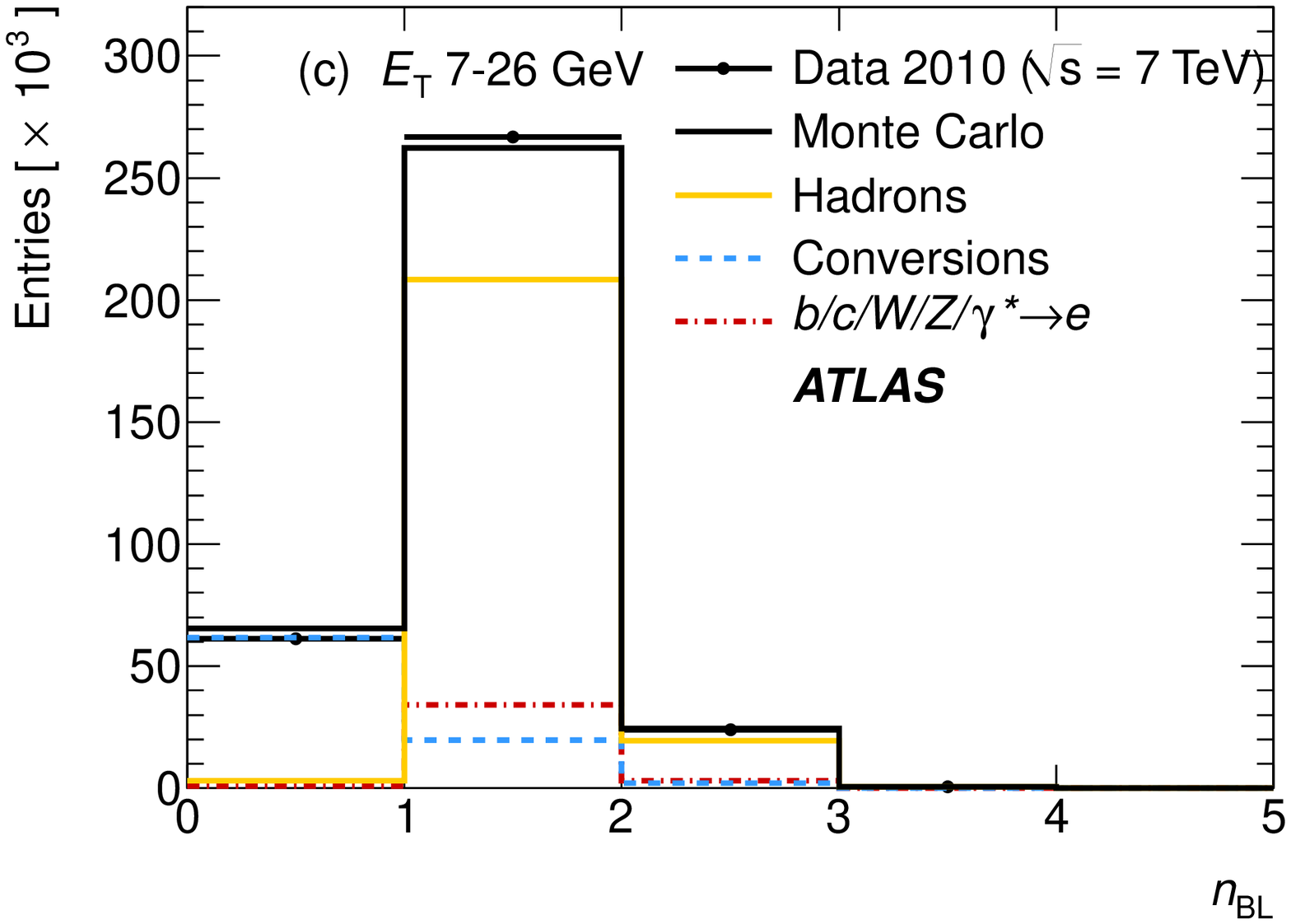} &
\includegraphics[width=0.48\textwidth]{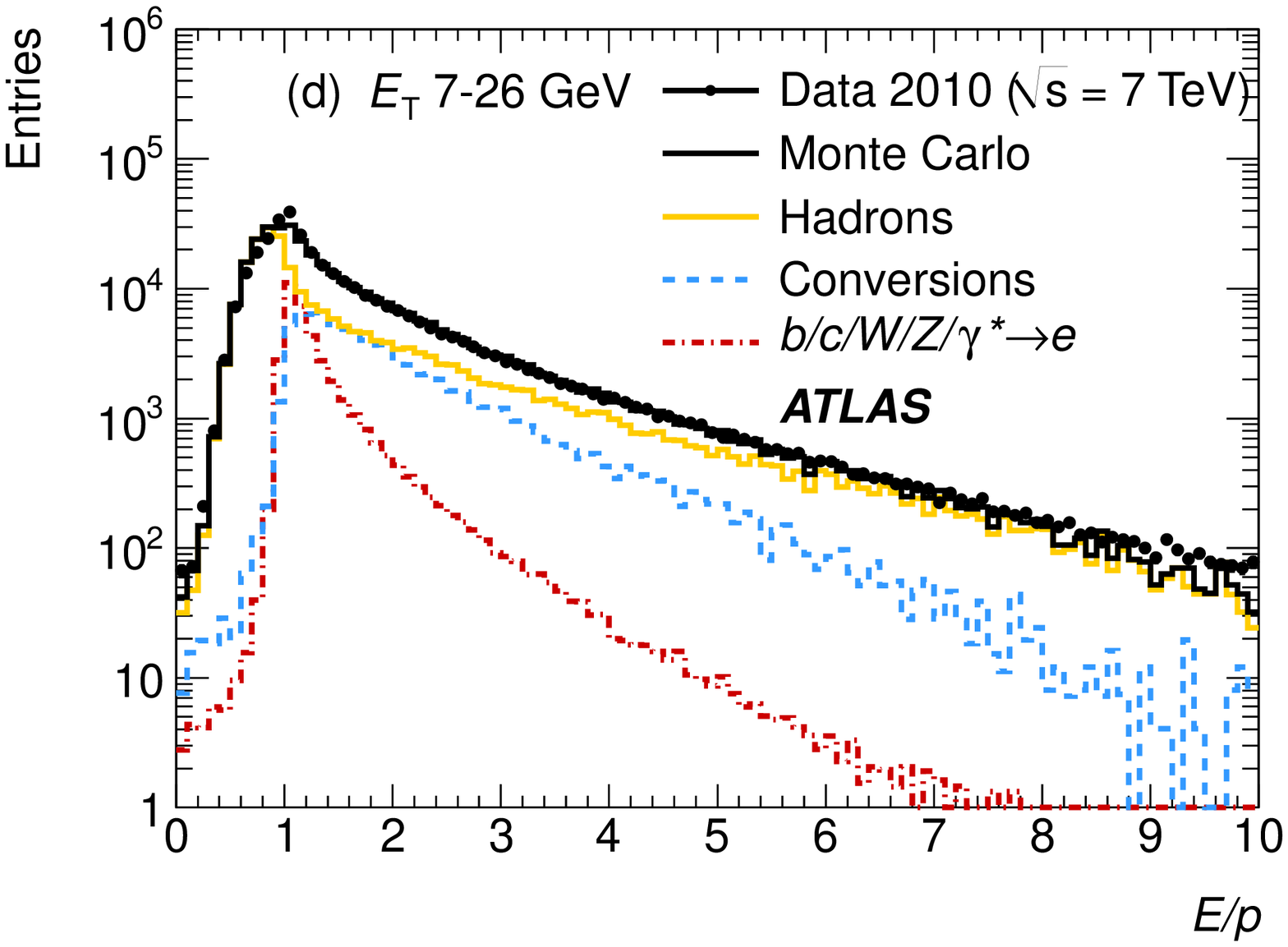}\\
\end{tabular}
\end{center}
\caption{\it\small (a) Distribution of cluster transverse energy, \et,~for the electron candidates.
 The simulation uses {\tt PYTHIA} with the $W$ and $Z/\gamma^{\ast}$ components 
normalised to their NNLO total cross-sections and the heavy-flavour, conversion and hadronic components  
then normalised to the total expectation from the data.
(b-d) {\tt PYTHIA} simulations of the distributions of discriminating variables used to extract the 
electron heavy-flavour plus $W$/$Z$/$\gamma^{\ast}$ signal compared to data:
(b) the 
ratio, $f_{\rm TR}$, between the number of high-threshold hits and all TRT hits on the electron track; (c) 
the number of hits,~$n_{{\rm BL}}$, on the electron track in the pixel $B$-layer; (d) 
the ratio,~$E/p$, between cluster energy and track momentum. 
}
\label{fig:ElectronKinematics}
\end{figure}

 The cluster transverse energy spectrum for the selected electron candidates in the data and simulation is shown 
in Fig.~\ref{fig:ElectronKinematics}(a), in which 
data with $\et < 18$~GeV have been rescaled to 1.3~pb$^{-1}$ from lower integrated luminosities. 
The discontinuities in the spectrum at 10, 20 and 30 GeV correspond to boundaries in the \et-dependent 
identification cuts, which mainly affect the yield of hadronic fakes. 
The  candidates in the simulation are sub-divided according to their dominant origins,
which for $\et<26$~GeV are non-isolated signal electrons from semi-leptonic decays of charm and beauty hadrons ($\sim$10\%),
a background of secondary electrons, largely dominated by electrons from photon conversions  ($\sim$20\%) and the 
dominant background of misidentified hadronic fakes.
 The fraction of isolated signal electrons from $W$/$Z$/$\gamma^{\ast}$ production is also shown. 
For $\et>26$~GeV this contribution starts to become significant, with the efficiency of
the identification cuts being higher for these isolated electrons, motivating the 
choice of the restricted 7-26~GeV analysis region.

The signal purity could be improved through the application of further cuts on
the fraction, $f_{\rm TR}$, of high-threshold (transition radiation) TRT hits out of all TRT hits measured on the 
track, the number of hits in the pixel B-layer, $n_{{\rm {BL}}}$ and 
the ratio of the measured energy of the EM cluster to the track momentum, $E/p$.
These variables offer excellent discriminating power against the hadronic fake ($f_{\rm TR}$) 
and photon conversion ($n_{{\rm {BL}}}$) backgrounds, as illustrated in 
Fig.~\ref{fig:ElectronKinematics}(b-d).
Applying tighter cuts on these variables would
increase the signal fraction to 50\% in the range 7 $< \et <$ 26 GeV but leave
 no means of estimating the remaining background fraction from data.
The cuts are therefore not applied, permitting the full distributions of the
variables to be used in the fitting procedure
described in Section 5.3.

\subsection{Electron trigger efficiency measurement}

The efficiency with which the signal electrons pass the L1 EM trigger is measured from the data in bins of
cluster \et .
For the 3 and 6~GeV threshold triggers, the efficiencies are measured using events selected by an alternative, 
very inclusive minimum bias
trigger, based on hit information in the Minimum Bias Trigger Scintillator~\cite{MBTS}.
The efficiencies of the 11 and 15~GeV triggers are measured using events recorded by the
6~GeV trigger, which is fully efficient in the \et~region for which
the higher threshold triggers are used.
Since these data-derived measurements are performed on the selected electron candidates, dominated by the hadronic background, 
a systematic uncertainty is estimated by comparing the
measured trigger efficiencies to those expected in the simulation for heavy-flavour electrons.
The trigger efficiencies are measured to be between 92.1\% and 100.0\%, with a maximum uncertainty of 1.8\%.

\subsection{Electron signal extraction}

In order to extract the heavy-flavour plus $W$/$Z$/$\gamma^{\ast}$ signal electrons from the selected candidates,
 a binned maximum likelihood method is used, based on the distributions of $f_{\rm TR}$, $n_{{\rm {BL}}}$ and  $E/p$. 
From simulation, a twelve-bin three-dimensional  probability density function (pdf) in these variables is constructed 
for the signal and conversion components.
For the  hadronic background, the shapes of the three template distributions
are described by additional free parameters (as in~\cite{tiles}) and are fitted to the 
data:
 in doing so, the method assumes no 
correlations exist between the three discriminating variables in the hadronic component.
The likelihood fit is performed in bins of $\eta$ (on which the discriminating distributions depend) and in~\et~in 
the range 7-26~GeV, allowing the fraction of signal, conversion and
hadronic fake candidates to be found in each~\et~bin.

The systematic uncertainty on the number of extracted signal electrons arising from
the differences between the data and simulation in the discriminating variables for the signal and conversion components is estimated to be less than 4\%, evaluated by repeating the signal extraction with the signal and conversion 
templates adjusted within their systematic uncertainties.
For $f_{\rm TR}$ and $E/p$, which have the largest effect, the differences were evaluated by comparing the
distributions in data and simulation for a pure sample of photon conversions, selected
by imposing the additional requirements of $n_{{\rm {BL}}}=0$ and either $E/p > 0.8$ or $f_{\rm TR}>0.1$, respectively.
The impact of the finite statistics of the simulated samples ($<$~2.5\%) and any possible bias in the method (7.3\%)
arising from the assumption that the template distributions for the hadron background are uncorrelated were
studied using pseudo-experiment techniques. 
The uncertainty associated with the electron energy scale (3.5\%) has been assessed by varying the electron candidate cluster 
energy by 1\% for $|\eta|<1.4$ and by 3\% for $|\eta|>1.4$, these systematic effects having been evaluated from 
$Z\rightarrow e^+e^-$ events.
Overall a statistical (systematic) uncertainty on the extracted signal component of approximately 3 (9)~\% is obtained.

\subsection{Determination of the electron efficiency and migration correction}

The overall efficiency and migration correction factor, $\epsilon_{\mathrm{(reco+PID)}_i}/C_{{\rm {migration}}_i}$, 
is determined from {\tt PYTHIA}-simulated samples of heavy-flavour decays to electrons and varies between 0.6 and 0.7 as a function of the
true electron \pt. 
Efficiencies of individual cuts were cross-checked on data control samples where possible, 
and a systematic uncertainty of 5-10\% is estimated  by recalculating
$\epsilon_{\mathrm{(reco+PID)}_i}$ $/C_{{\rm {migration}}_i}$ from simulated samples produced
with an increase in the amount of material inside the EM calorimeter,
corresponding to the estimated uncertainty on the material budget (see~\cite{egammapaper} and references therein). The statistical uncertainty on $\epsilon_{\mathrm{(reco+PID)}_i}/C_{{\rm {migration}}_i}$ is found to be between 0.4 and 3.5\%.

Additionally, the efficiency of the electron identification cuts in the simulation is compared with a measurement 
made on data using a tag-and-probe (T\&P) technique.
The probe candidates, which must pass only the preselection
cuts of Section~5.1, are taken from a sample of 
events enriched in heavy quark pair production 
where both  heavy hadrons decay semi-leptonically. 
To select such events, the tag electron candidate is subject to more stringent identification cuts
than those described in~5.1, 
including requirements on $f_{\rm TR}$ and $n_{\rm {BL}}$, and the T\&P candidate
pair must have opposite charge and
an invariant mass below the $Z$ mass window and outside of the $J/\psi$ mass region.
The signal purity remains low after the T\&P selection, being 9 (31)~\% for probe candidates before (after) applying the identification criteria.
The signal component of the probe candidates before and after  the identification cuts must therefore be extracted with a method similar to that
described in Section 5.3.  
By comparing the measured identification efficiency of the extracted probe electrons to that expected in simulation as a function of \et, an uncertainty of 5\% is obtained on the identification efficiency, with a further 7\% systematic uncertainty coming from the T\&P method itself. 

Overall the uncertainty on $\epsilon_{\mathrm{(reco+PID)}_i}/C_{{\rm {migration}}_i}$ is found to be 12-14\%, depending on the true 
electron \pt.
Possible effects of the choice of heavy-flavour hadron decay model and the prompt  J/$\psi$ contamination are found to be negligible.

\subsection{Electron production cross-section result} \label{sec:ele_inclusive}

The differential cross-section for electrons from heavy-flavour production
is found from Eqn.~1 using a bin-by-bin unfolding method.
Before applying the efficiency and migration correction factor, $\epsilon_{\mathrm{(reco+PID)}_i}/C_{\mathrm{migration}_i}$,
the theoretical prediction for the accepted electron cross-section from  $W$/$Z$/$\gamma^{\ast}$ decays,
$\sigma^\mathrm{W/Z/\gamma^*}_\mathrm{accepted}$,  must first be subtracted.\footnote{The uncertainty on the heavy flavour cross-section arising from the overall
uncertainty on $\sigma^\mathrm{W/Z/\gamma^*}_\mathrm{accepted}$ 
is negligible, reaching at most 1\% in the highest \pt~bin where the  $W$/$Z$/$\gamma^{\ast}$ contribution to the signal reaches its maximum of 13\%.}
$\sigma^\mathrm{W/Z/\gamma^*}_\mathrm{accepted}$ is obtained
from {\tt PYTHIA}, with the high-mass $W$/$Z$ contribution normalised to the NNLO
total cross-section~\cite{fewz,ATLAS_W_Z}

The differential cross-section for electrons from heavy-flavour production
within $|\eta| < 2.0$ (excluding  $1.37 < |\eta| < 1.52$) and  $7<{\pt}<26$~GeV
is plotted in Fig.~\ref{fig:FullComparison} (left) and reported in Table~\ref{tab:crosssection}.
The statistical uncertainty originates from the signal extraction procedure
(Section 5.3), and the sources of systematic uncertainty, as discussed
in the preceding sub-sections, are summarised in Table~\ref{tab:Summary}. 
Correlations between the systematic uncertainties common to the signal extraction and the T\&P efficiency measurement, 
such as discrepancies between the data and simulation in the signal and conversion pdfs and the energy scale
uncertainty, are taken into account in the evaluation of the overall 
systematic uncertainty on the cross-section.
To account for possible biases due to the \pt~distribution of the signal,
the predictions of simulated heavy-flavour samples from different programs 
({\tt {PYTHIA}}, {\tt {POWHEG+PYTHIA}} and {\tt {POWHEG+HERWIG}})
are compared and found to yield consistent  results.

We obtain a fiducial heavy-flavour electron cross-section  
in the range $ 7 < \pt < 26$~GeV and within $|\eta| <2.0$, excluding $1.37<|\eta|<1.52$, of
\begin{equation*}
\sigma^e_{\rm HF} = 0.946 \pm 0.020{\rm (stat.)} \pm 0.146{\rm (syst.)} \pm 0.032{\rm (lumi.)}~\mu \rm{b}.
\end{equation*}

\begin{table}[htb!]
\caption{\it\small 
Summary of systematic uncertainties on the electron heavy-flavour cross-section. 
The uncertainties apply in the \pt~bins of the measurement; an interval or upper limit is given where the 
uncertainty varies as a function of~\pt. 
Correlations between the systematic uncertainties reported independently for the signal extraction and the T\&P efficiency measurement
in Sections 5.3 and 5.4 are taken into account.
}
\begin{center}
\footnotesize
\begin{tabular}{l|l}
\hline
\multirow{2}{*}{Source of systematic uncertainty}	& Cross-section \\
                                        & uncertainty (\%) 	\\
\hline
\hline
Energy scale uncertainty    	& 1.5		\\
\hline
Possible bias in signal extraction      & 8 \\
Mis-modelling of discriminating variables & 8	\\
Stat. uncertainty on pdfs for
signal extraction			& 0.8$-$2.5 \\
\hline
Material uncertainty on 
$\epsilon_{\mathrm{(reco+PID)}_i}/
C_{\mathrm{migration}_i}$ 			& 5$-$10\\
Stat. uncertainty on 
$\epsilon_{\mathrm{(reco+PID)}_i}/
C_{\mathrm{migration}_i}$			& 0.4$-$3.5	\\
Efficiency dependence on \pt\ from T\&P	        & 5	\\
\hline
Trigger efficiency (stat.+syst.)		& $<2$\\
\hline
Accepted $W$/$Z$/$\gamma^{\ast}$ cross-section 
(stat.+syst.)				& $<1$\\
\hline
Integrated luminosity 				& 3.4\% 	\\
\hline
\hline
Total & 14$-$17 \\
\hline
\end{tabular}
\end{center}
\label{tab:Summary}
\end{table}

\section{Muon analysis}

\subsection{Muon candidate selection}
Muon candidates within a pseudorapidity of $|\eta| < 2.5$ are 
selected if they have at least two MDT segments and an ID track with hits in two different sub-detectors.
In addition to signal muons from charm, beauty and  $W$/$Z$/$\gamma^{\ast}$ decays, the selected candidates
comprise a significant fraction of background  muons
from pion and kaon decays in flight ($\pi/K$) and misidentified muons from  hadronic showers in the calorimeter that reach the MS and are wrongly matched to
a reconstructed ID track (fakes).
The  $\pi/K$  background is subdivided into those that decay close enough to the IP such that the
majority of hits on the ID track come from the decay muon (early-$\pi/K$) and those that do not (late-$\pi/K$).
The signal purity of the sample, determined using the method discussed in Section 6.4, ranges from
45\% at $\pt = 4$~GeV to 90\% at 40~GeV in the region of the $W$/$Z$
Jacobian peak.

\subsection{Muon trigger  efficiency measurement}
The trigger efficiency for the muon candidates is evaluated 
using events recorded by an independent trigger 
based on calorimeter information alone. 
 The efficiency for the lower threshold trigger is found to be 68\% 
at \pt~$=$~4 GeV and to reach a plateau of 84\% at 9 GeV. The 10 GeV threshold trigger
efficiency is constant for \pt~$ > 16$ GeV with a value of 74\%. (The muon trigger efficiency is dominated by the limited acceptance of the muon trigger chambers). 
The data samples used to compute the efficiency contain background muons. In order to obtain
the efficiencies for signal muons, correction factors of 1.04 for the low threshold~trigger and 1.08 for
the 10 GeV trigger   are estimated from  
simulation.  Systematic uncertainties on these correction factors come
from the simulation statistics (0.5\% and 0.7\% for the lower threshold and 10~GeV triggers, respectively) and  from the
mis-modelling of the signal fraction by the simulation (0.7\% and 0.2\% for the two triggers), the
latter being assessed by reweighting the 
simulated sample according to the measured signal fraction.
Other sources of systematic uncertainty arise from the statistical
fluctuations in the independent trigger sample (from 0.4\% to 0.9\% for the low threshold trigger, and 0.5\% for the 10 GeV trigger)
and from the bias
introduced by the independent trigger (evaluated to be 2.3\% for the 10 GeV trigger by
comparing to events triggered by the low threshold trigger).

\subsection{Muon reconstruction efficiency measurement}

The combined muon reconstruction efficiency has 
three components:
the ID efficiency ($\epsilon_{\rm ID}$), the MS
efficiency ($\epsilon_{\rm MS}$) and the matching efficiency
($\epsilon_{\rm Match}$).
The overall efficiency has been determined from high-statistics simulated muon samples from 
heavy-flavour hadron and $W$/$Z/\gamma^{\ast}$ decays,
with correction factors for each component of the reconstruction,  $\alpha_{x} =
\epsilon_{x}^{\rm data}/\epsilon_{x}^{\rm simulation}$  ($x =
$~ID, MS, Match),  being determined by comparing the 
simulation-derived efficiencies with those observed in data.
The overall reconstruction efficiency is 
found to be 85\% at $\pt=4$~GeV,
reaching 95\% at 7 GeV. The plateau value of 95\% is the same for both isolated and non-isolated
muons.

The ID correction factor $\alpha_{\rm ID}$ is evaluated with a T\&P
method on $J/\psi$ and $Z$ events, using a combined muon track 
as a tag and an MS track as a probe. The
fraction of ID tracks found over the number of probes has
been computed and compared to the expectation in simulation, giving a value of 
$\alpha_{\rm ID} = 1.000 \pm 0.005$, where the quoted uncertainty includes both the statistical
and systematic contributions.

The product $\alpha_{\rm MS} \cdot
\alpha_{\rm Match}$ is obtained with two methods.
 The first method identifies single muon
tracks  in jets from energy deposits
corresponding to minimum-ionising particles
in calorimeter cells matched to extrapolated ID tracks.
In order to reject the background from pions and kaons from the primary
vertex, a cut on the impact parameter ($d$) to the primary vertex in the
transverse plane is applied:  $|d/\sigma_{d}| > 3$, where
$\sigma_{d}$ is the error on $d$ from the tracking
algorithm. According to simulation this cut selects muons from beauty decays with a purity of
99\%. The factor $\epsilon^{\rm data}_{\rm MS}\cdot \epsilon^{\rm data}_{\rm Match}$ is then computed by evaluating
the fraction of these tracks that are reconstructed in the MS and matched to the ID track.
  The second method identifies muons by matching  ID tracks with hits in the MS trigger chambers. 
The trigger bias of this method has been evaluated with simulated data to be 2\% for $p_{\rm T} <  6$ GeV and less than 0.2\% at higher momenta. 
Overall a value of  $\alpha_{\rm MS} \cdot \alpha_{\rm Match} = 0.986 \pm 0.003{\rm (stat.)} \pm 0.010 {\rm (syst.)}$ 
is obtained, the central value being the average of the results from the two
methods and the systematic uncertainty coming from the difference between the
two. Both methods are sensitive up to $\pt = 30$ GeV, in the region where
 the control sample is dominated by the non-isolated muons.
To take into account isolated muons and muons with $\pt > 30 $~GeV, 
the result has been compared with  $\alpha_{\rm MS} \cdot \alpha_{\rm Match}$ 
computed from two other T\&P techniques, using muons from J/$\psi$ \cite{JPsi_CONF} and $Z$ \cite{Z_CONF} decays.
The T\&P technique used here is the same as that used
in the determination of $\epsilon_{\rm ID}$ but with the probe muon selected among
the ID tracks, and a full combined track being required in the numerator.
The  $\alpha_{\rm MS} \cdot \alpha_{\rm Match}$ scale factors obtained with the T\&P methods are 
fully compatible with those obtained using the single muon track methods. 

Overall the systematic uncertainty on the muon reconstruction efficiency is dominated by the uncertainties on the scale factors reported
above and evaluates to 1.2\%. 

\subsection{Muon signal extraction}

\begin{figure}[h]
\begin{center}
\includegraphics[width=0.325\textwidth]{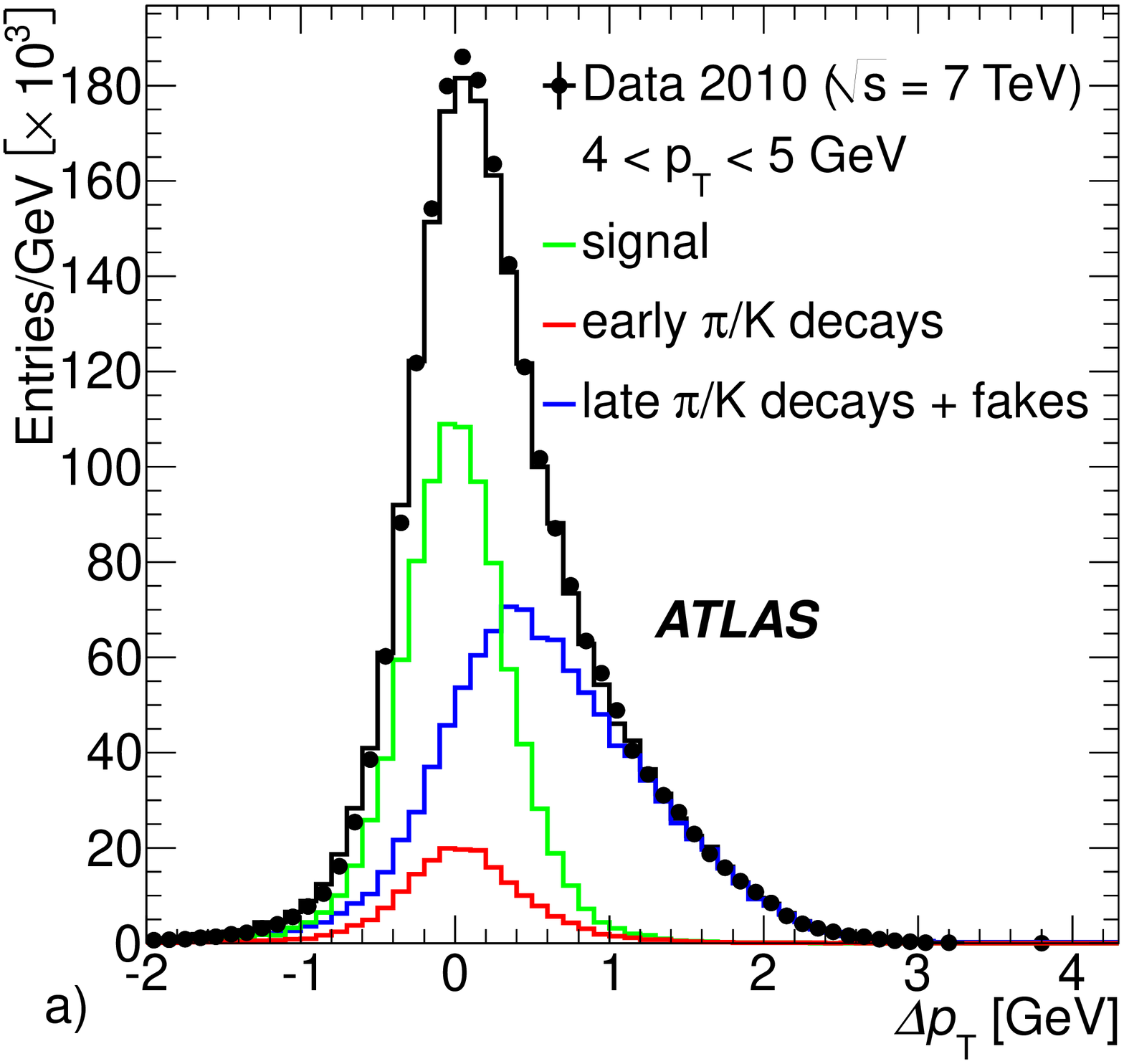}
\includegraphics[width=0.325\textwidth]{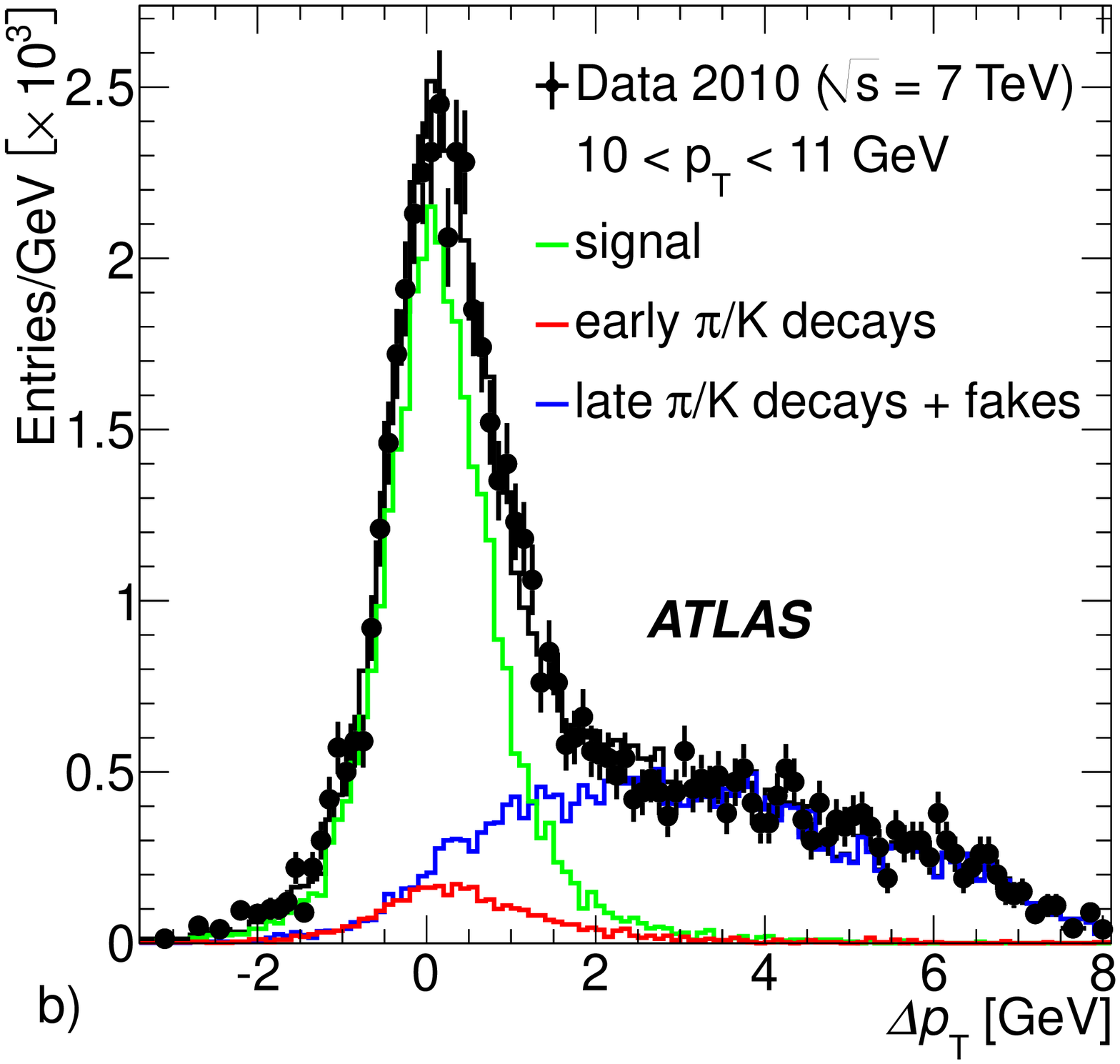}
\includegraphics[width=0.325\textwidth]{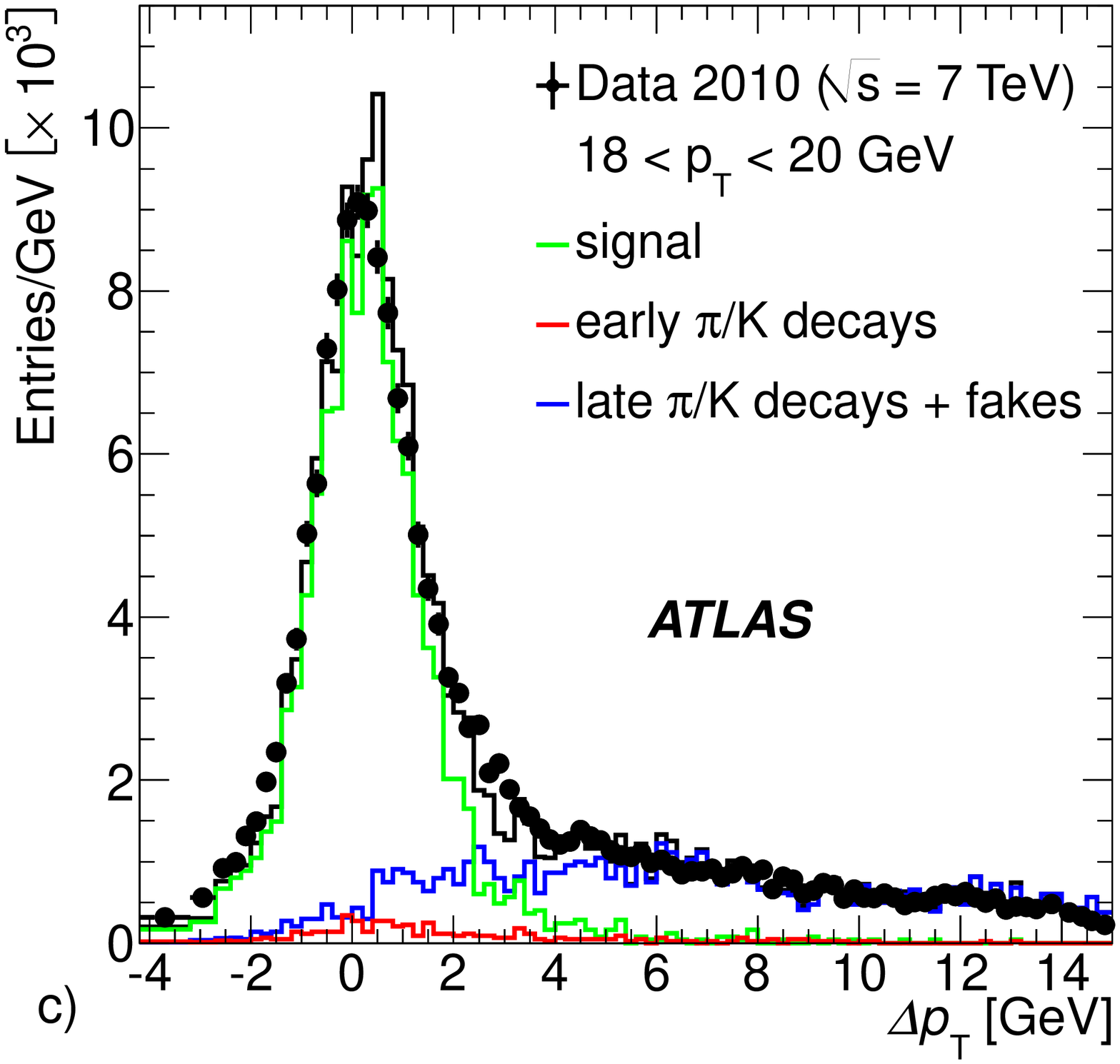}

\caption{\it \small The $\Delta \pt$ distribution in the \pt~bins 4-5 GeV
  (a),  10-11 GeV (b)
  and 18-20 GeV (c) for muon combined track candidates. The signal, 
  early-$\pi/K$ and late-$\pi/K$ plus fakes components from simulation are shown.
The early-$\pi/K$s are defined as those that decay close enough to the IP that the majority of hits on the ID track come from the decay muon.}
\label{fig:DeltaPt}
\end{center}
\end{figure}

The muon reconstruction provides independent information on the $\pt$ of the track reconstructed in the ID
 and in the MS. 
The difference in \pt , $\Delta \pt = \pt^{\rm ID} - \pt^{\rm MS}$, where both momenta are extrapolated 
to the IP, is sensitive to the origin of the muons: signal, early-$\pi/K$, and late-$\pi/K$ or fakes, 
as illustrated in Fig.~\ref{fig:DeltaPt} for three $p_{T}$ intervals. A fit to the data distribution is performed to extract the signal component using templates from the simulation.  
The  early-$\pi/K$ component template, like the signal, has a 
 $\Delta \pt$ distribution  peaked around zero,
since the \pt~reconstructed in the ID for a $\pi/K$ that decays close to the IP
is dominated by hits from the decay muon.
The late-$\pi/K$ component
and the fake component may be described by a single template with a  broader
 $\Delta p_{T}$ distribution shifted towards higher values. 
Since the early-$\pi/K$ component is significant only for $\pt<10$~GeV and cannot be strongly discriminated from the signal, 
we fix the ratio of the early-$\pi/K$ component to the late-$\pi/K$ plus fakes component to its expectation in the simulation
and use only a single background template in the fit.
A systematic uncertainty is assigned to cover the possible
difference in the (early-$\pi/K$)/(late-$\pi/K$+fakes) ratio between data and simulation as explained below.

The fit is performed in $\pt$ bins over the whole range. For $\pt < 52$~GeV the template distributions are taken from
a {\tt PYTHIA} dijet sample with $\hat{p}_{\rm T} >
15$ GeV (where $\hat{p}_{\rm T}$ is the \pt~of the primary parton) 
with the additional requirement that  at least one set of particles crossed
a surface of
$\Delta \eta \times \Delta \phi = 0.12 \times 0.12$ with a total energy greater than 17
GeV. For \pt~$> 52$ GeV a dijet sample with $\hat{p}_{\rm T} > 280$ GeV
is used. 

The systematic uncertainty on the extracted signal fraction arising from the difference in
the $\Delta \pt$ distributions between the simulated template samples and the
expected data distributions is evaluated on simulated samples of QCD jets (light and heavy-flavour) 
and $W$/$Z$ inclusive events 
that reproduce the expected composition of data. The maximum possible bias is found to be 3\%.
The effect of any mis-modelling of the background $\Delta p_{\rm{T}}$ template is also checked by comparing
the extracted signal fraction to that obtained when using 
a  background template 
taken from a simulated sample whose
$\pt$ spectrum is weighted to reproduce the spectrum observed in data
before the signal extraction.
A difference of 1.5\% is found, within the bias mentionned above. Therefore we quote an overall 3\% systematic uncertainty for the template modelisation. 
The  systematic uncertainty on the signal fraction due to the finite statistics of the simulated samples 
used for the template distributions is found to be between  1\% and 8\%.

The accuracy of the assumption that the ratio of the early-$\pi/K$ component
to the combined late-$\pi/K$ plus fakes component, $r$, is reproduced correctly by the simulation is tested by comparing
the $\Delta \pt$  distributions in data and simulation as a function of the early-$\pi/K$ fraction.
A correction factor $r_{\rm data}/r_{\rm simulation}$ is determined as 1.1 $\pm$ 0.1. This 10\% uncertainty on
 $r$ corresponds to an uncertainty on the signal of 2\% at \pt~$=$~4 GeV, rapidly falling to zero for \pt~$> 10$ GeV.

\subsection{Muon resolution and unfolding}\label{sec:unfolding}

The muon momentum resolution has been studied using tracks from the
decays $Z \to \mu^{+} \mu^{-}$ and $J/\psi \to \mu^{+} \mu^{-}$.
With an iterative procedure, the simulated muon track momenta  are smeared and scaled as a
function of pseudorapidity to  reproduce the $J/\psi$ and
the $Z$ invariant mass shapes measured in data~\cite{ResolutionMuons}.
A full set of smearing parameters for the MS and ID are obtained, and the 
corresponding effect on the combined muon derived.
The corrected sample is used to obtain the unfolding coefficients
$C_{\mathrm{migration}_i}$ in Eqn.~\ref{eq:cross_section}.
The uncertainty on the unfolding coefficients is determined by
varying independently the cross-section values of the heavy-flavour 
and $W$/$Z$ components by 30\% and 10\% respectively. 
The associated systematic uncertainty is at the level of 0.1\%
over almost the whole spectrum with a maximum value of 1.2\% around the $W/Z$
Jacobian peak.

\subsection{Muon production cross-section result} \label{sec:muon_inclusive}

\begin{figure}[t!]
\centerline{\epsfig{file=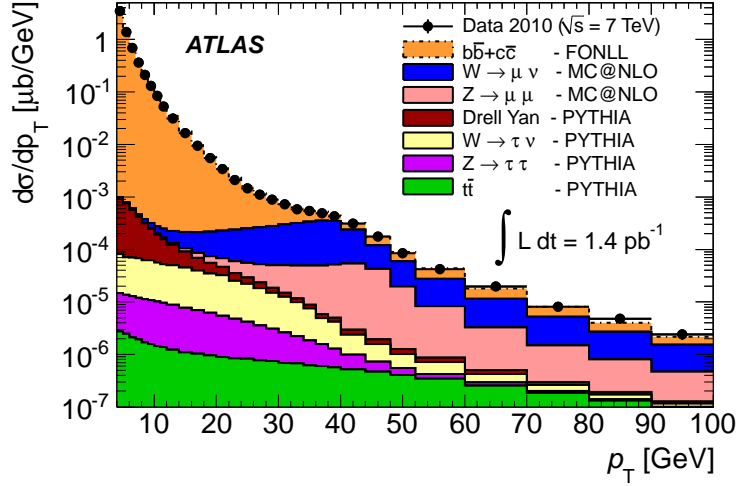, height=6.8cm}}
\caption{\it\small  Muon differential cross-section as a function of
  the muon transverse momentum for  $|\eta| <2.5$ compared to theoretical predictions. The Drell-Yan component corresponds to the $Z$/$\gamma^{\ast}$ for M$_{\mu^+ \mu^-} < 60$ GeV.}
\label{fig:MuonPart}
\end{figure}

The signal fraction of the muon transverse momentum spectrum has been corrected for the trigger and
reconstruction efficiencies and unfolded from the detector
response. Figure~\ref{fig:MuonPart} shows the resulting inclusive muon differential
cross-section for muons within $|\eta| < 2.5$ as a function of \pt, compared to the overall theoretical
expectation.
 The  expected $W$/$Z$ component comes from
{\tt MC@NLO}~\cite{mcatnlo1, mcatnlo2} using the {\tt CTEQ6.6} PDFs,  normalised to the cross-sections for muons measured by ATLAS~\cite{ATLAS_W_Z}{\footnote {$\sigma^{\mu}_{W^+} = 6.21 \pm 0.02{\rm (stat.)} \pm 0.25{\rm (syst.)}$~nb, $\sigma^{\mu}_{W^-} = 4.107 \pm 0.02{\rm (stat.)} \pm 0.19{\rm (syst.)}$~nb, $\sigma^{\mu}_{Z} = 0.941 \pm 0.008{\rm (stat.)} \pm 0.038{\rm (syst.)}$~nb, where the systematic uncertainty excludes contributions from the luminosity and acceptance which are fully correlated with those presented here.}}. The FONLL prediction is used for the heavy-flavour component and the remaining, small contributions are obtained from {\tt PYTHIA} simulation. 
 The theoretical uncertainty is dominated by the heavy-flavour prediction, being approximately 20\% and discussed in Section 7, and is not shown in the figure. 
The systematic uncertainties on the measurement are summarised in Table~\ref{tab:Summary_muons}.

\begin{table}[h]
\caption{\it\small Summary of systematic uncertainties on the muon cross-section measurement.
The uncertainties apply in the \pt~bins of the measurement; an interval or upper limit is given where the 
uncertainty varies as a function of~\pt.}
\begin{center}
\footnotesize
\begin{tabular}{l|l}
\hline
                                         &  Cross-section \\
Source of systematic uncertainty         &  uncertainty \\
                                         & (\%)  \\
\hline 
\hline
Possible bias in signal extraction & 3 \\ 
Early-$\pi/K$ fraction & $<$2 \\
Stat. uncertainty on  signal extraction templates & 1$-$8 \\
\hline
Efficiency scale factor & 1.2 \\
\hline
Trigger efficiency  control sample statistics & 0.4$-$0.9 \\
Trigger efficiency  control sample bias & $<$2.3 \\
Trigger efficiency  background bias & 0.2$-$0.7 \\
Trigger efficiency  mis-modelling of signal fraction  & 0.5$-$0.7 \\
\hline
Unfolding procedure & 0.1$-$1.2 \\
\hline
Integrated luminosity                      & 3.4 \\
\hline
\hline
Total  &  5$-$8\\
\hline
\end{tabular}
\end{center}
\label{tab:Summary_muons}
\end{table}

 Integrating over the full 4-100~GeV \pt~range, in $|\eta| < 2.5$, we find a fiducial cross-section for inclusive muons of
\begin{equation*}
\sigma^{\mu}_{\rm Inc.} = 6.55 \pm 0.01{\rm (stat.)} \pm 0.37{\rm (syst.)} \pm 0.22{\rm (lumi.)} ~\mu {\rm b}.
\end{equation*}
In order to compare to the results of the electron analysis, the muon cross-section
has been studied in the same acceptance region ($7<{\pt}<26$~GeV and $|\eta| < 2.0$, excluding  $1.37 < |\eta| < 1.52$)
and with the subtraction of the $W$/$Z$/$\gamma^{\ast}$ contribution, giving a fiducial heavy-flavour muon cross-section  of 
\begin{equation*}
\sigma^{\mu}_{\rm HF} = 0.818 \pm 0.003{\rm (stat.)}  \pm 0.036{\rm (syst.)} \pm 0.028{\rm (lumi.)} ~\mu {\rm b}.
\end{equation*}

\section{Comparison of electron and muon cross-sections with theoretical predictions for heavy-flavour production}

\begin{figure}[ht!]
\includegraphics[width=0.5\textwidth]{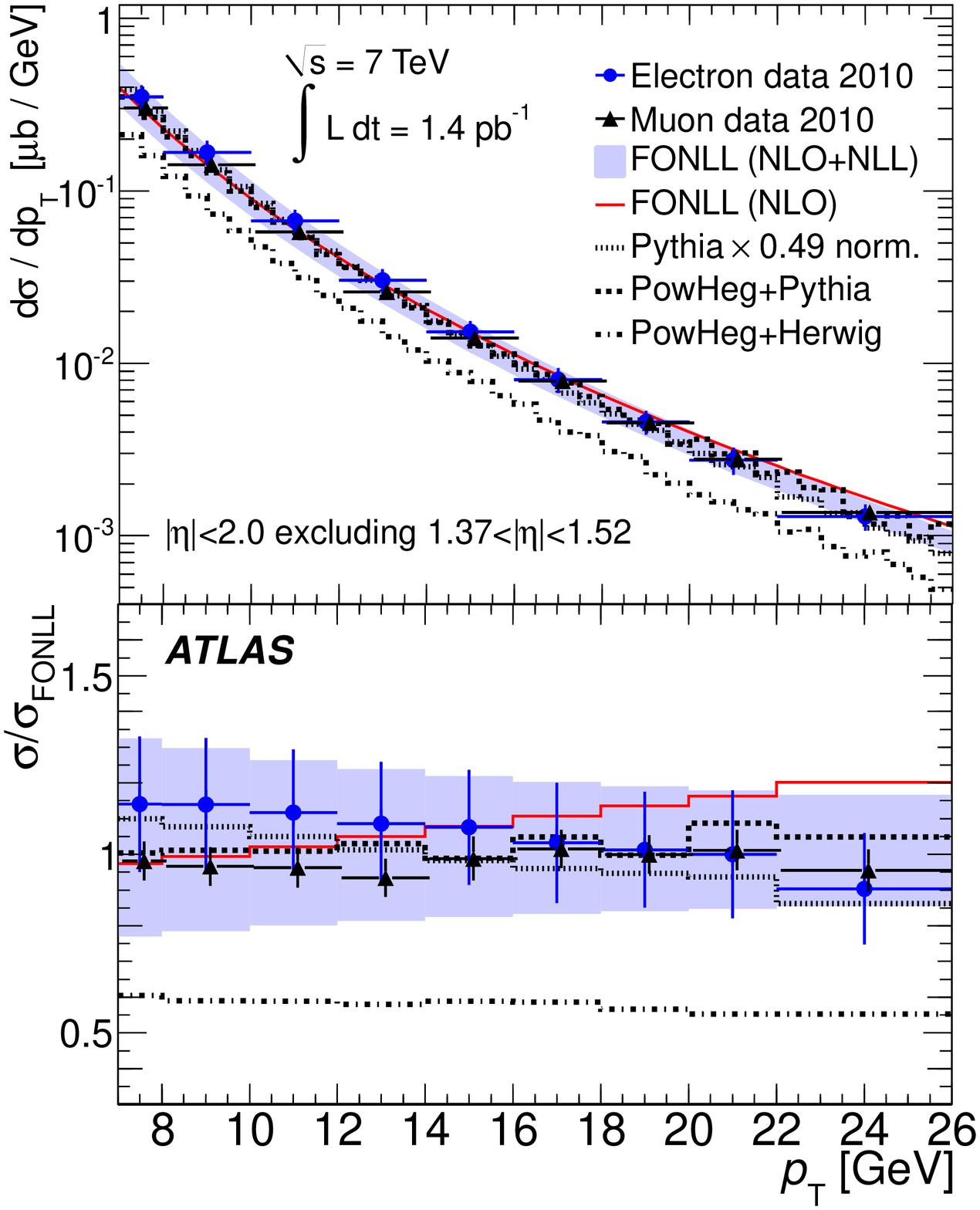}
\includegraphics[width=0.5\textwidth]{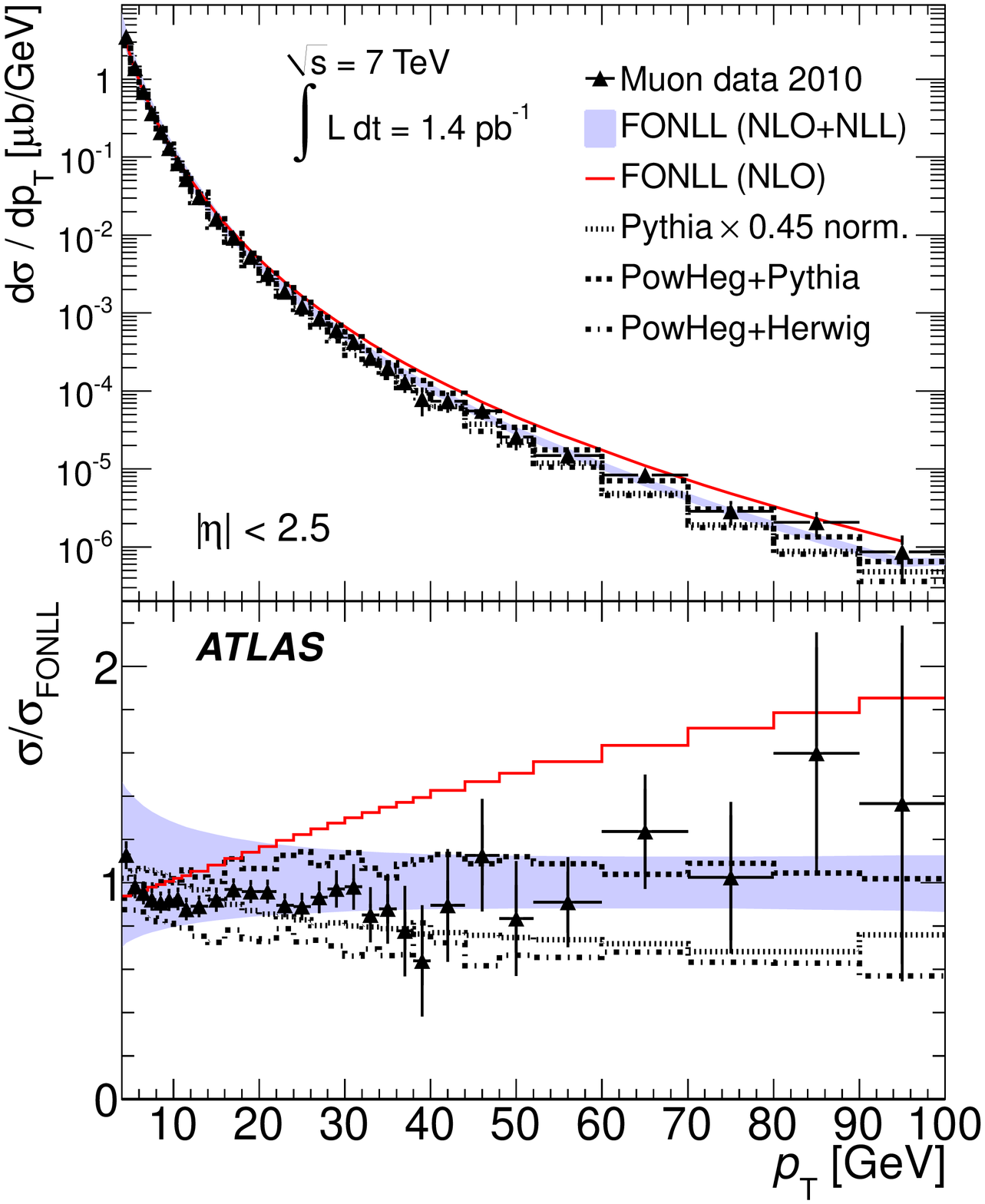}
\caption{\it \small
(Left) Electron and muon differential cross-sections from heavy-flavour production as a function of
the charged lepton transverse momentum for $|\eta| <2.0$  excluding the $1.37 < |\eta| <1.52$ region.
(Right) Muon differential cross-section as a function of the muon transverse momentum for  $|\eta| <2.5$. 
The data points include statistical and systematic uncertainties.
The ratio of the measured cross-section and the other predicted cross-sections to the FONLL calculation is given in the  bottom of each plot. 
The {\tt PYTHIA} (L0) cross-sections are normalised to the data in order to compare the shape of the spectra.
}
\label{fig:FullComparison}
\end{figure}

The results for both channels within the acceptance of the electron analysis (7 $<$ \pt~$ <$ 26 GeV,  
$|\eta|<2.0$ excluding $1.37<|\eta|<1.52$) are shown in 
Fig.~\ref{fig:FullComparison} (left) and 
summarised in Table~\ref{tab:crosssection}.
Additionally, the measured muon cross-section is given 
over the full \pt~(4-100 GeV)  and pseudorapidity ($|\eta| <
2.5$)  range in Fig.~\ref{fig:FullComparison} (right) and Table~\ref{tab:mucrosssection}.\footnote
{The results as shown in Tables~\ref{tab:crosssection} and~\ref{tab:mucrosssection} are 
available in the HEPDATA database~\cite{HEPDATA} and a Rivet~\cite{Rivet} routine is provided.}

In the electron analysis the $W$/$Z$/$\gamma^{\ast}$ contribution has been subtracted as
described in  Section \ref{sec:ele_inclusive} in order to obtain the heavy-flavour differential cross-section;
whereas, in the muon case the
subtraction is made at the cross-section level from the fully inclusive measurement.
The systematic uncertainties
on the electron and muon results are independent except for that arising from the uncertainty on the total luminosity. 
In the common acceptance region, the two measurements are in good agreement with each other.

\begin{table}[hbtp]
\newcommand\T{\rule{0pt}{2.6ex}}
\newcommand\B{\rule[-1.6ex]{0pt}{0pt}}
\caption{\it \small Differential cross-sections  $d\sigma/dp_{\rm T}{}_{e(\mu)}$ (in nb/GeV) for electron (muon) heavy-flavour production in the pseudorapidity region  $|\eta|<2.0$ (excluding $1.37<|\eta|<1.52$), with statistical (stat.) and systematic (syst.) uncertainties. 
The 3.4\% luminosity uncertainty is included in the latter.
The predictions of FONLL are also given.}
\setlength{\tabcolsep}{1pt}
\scriptsize
\begin{center}
  \begin{tabular}{c|ccccc|ccccc|rl}
\hline 
\pt interval[GeV]  \T  \B &  $\frac {d\sigma}{dp_{\rm T}}{}_{e}$&$\pm$&stat.&$\pm$&syst. 
                                &   $\frac {d\sigma}{dp_{\rm T}}{}_{\mu}$&$\pm$&stat.&$\pm$&syst. 
                                &  $\frac {d\sigma}{dp_{\rm T}}\!$&$_{\mathrm {FONLL} } $\\
\hline
\hline 
\T 7--8 &   351 &$\pm$& 15 &$\pm$& 56              &  302 &$\pm$& 2 &$\pm$& 17           &  308&$^{+115}_{-74}$\\
\T 8--10 &   167 &$\pm$& 6 &$\pm$& 27              &  142 &$\pm$& 1 &$\pm$& 8             &  146&$^{+50}_{-33}$\\
\T 10--12 &   67 &$\pm$& 2 &$\pm$& 11              &  58.0 &$\pm$& 0.6 &$\pm$& 3.0     &   60&$^{+19}_{-13}$\\
\T 12--14 &   30.3 &$\pm$& 1.1 &$\pm$& 4.7      &  26.1 &$\pm$& 0.4 &$\pm$& 1.4     &   28&$^{+8}_{-5}$\\
\T 14--16 &   15.3 &$\pm$& 0.4 &$\pm$& 2.2      &  14.1 &$\pm$& 0.3 &$\pm$& 0.8      &  14&$^{+4}_{-3}$\\
\T 16--18 &   8.0 &$\pm$& 0.3 &$\pm$& 1.3         &  7.92 &$\pm$& 0.05 &$\pm$& 0.41  & 7.8&$^{+2.0}_{-1.5}$\\
\T 18--20 &   4.58 &$\pm$& 0.15 &$\pm$& 0.72  &  4.52 &$\pm$& 0.04 &$\pm$& 0.24  &  4.5&$^{+1.1}_{-0.8}$\\
\T 20--22 &   2.75 &$\pm$& 0.09 &$\pm$& 0.48  &  2.78 &$\pm$& 0.03 &$\pm$& 0.15  &  2.7&$^{+0.6}_{-0.4}$\\
\T \B 22--26 &   1.29 &$\pm$& 0.05 &$\pm$& 0.21  &  1.37 &$\pm$& 0.02 &$\pm$& 0.08  &  1.4&$^{+0.3}_{-0.2}$\\
\hline                                 
\end{tabular}
\end{center}
\label{tab:crosssection}
\end{table}

\begin{table}[htp!]
\caption{\small \it  Differential cross-sections  $d\sigma_{\mu}/dp_{\rm T}$ (in nb/GeV) 
in the pseudorapidity region  $|\eta|<2.5$, before and after subtraction of
  the $W$/$Z$/$\gamma^{\ast}$ component, 
with statistical (stat.) and systematic (syst.) uncertainties. 
The 3.4\% luminosity uncertainty is included in the latter.
The uncertainty on the $W$/$Z$/$\gamma^{\ast}$ component is not included and amounts to 4\% of the subtraction, increasing the systematic error by 5--10 \% for $p_{\rm T} > 32$~GeV.}
\begin{center}
\scriptsize
\begin{tabular}{c|c|c|c|c}
\hline
$\pt$ interval&\multirow{2}{*}{$d\sigma_{\mu} /dp_{\rm T}$}&$d\sigma_{\mu} /dp_{\rm T}$&\multirow{2}{*}{ $\pm$ stat.}&\multirow{2}{*}{$\pm$ syst.}\\
{[}GeV{]} &   & $W$/$Z$/$\gamma^{\ast}$ sub. & & \\
\hline
\hline 
4--5 & 3490 & 3490 & 7   &  230  \\
5--6& 1390 & 1390 & 4 & 90  \\
6--7& 680 & 680 & 3  &  40  \\
7--8& 364 & 364 & 2   & 23  \\
8--9& 210 & 210 & 2  & 13  \\
9--10& 130 & 130 & 1 & 8  \\
10--11 & 84 & 84 &  1 & 5  \\
11--12 & 53 & 53 & 0.9   & 3  \\
12--14 &  31 &  31 & 0.5   & 2  \\
14--16 & 16.3 & 16.1 & 0.4 & 1.1  \\
16--18 & 9.4 & 9.2  & 0.06  &  0.6  \\
18--20 & 5.5   & 5.3   & 0.04   & 0.3   \\ 
20--22 & 3.4 & 3.2  & 0.03 &  0.2  \\
22--24 & 2.11  & 1.87  & 0.03  &  0.13  \\ 
24--26 & 1.46  & 1.20  & 0.02  & 0.10  \\
26--28 & 1.11  & 0.84   &  0.02  &  0.07  \\
28--30 & 0.88   & 0.60  &  0.02   &  0.06    \\
30--32 & 0.73   & 0.43  & 0.02  &  0.05  \\
32--34 & 0.58   &  0.26  &  0.02   &  0.04  \\
34--36 & 0.54   &  0.20  &  0.02  &  0.03  \\
36--38 & 0.48   &  0.13  & 0.01   & 0.03  \\
38--40 & 0.428  &  0.088  &  0.014  &  0.029  \\
40--44 & 0.311  &  0.074  &  0.009  &  0.020   \\
44--48 & 0.176  &  0.056  &  0.007  &  0.011  \\
48--52 & 0.085  &  0.025  &  0.005  &  0.007   \\
52--60 & 0.042  &  0.015  &  0.002  &  0.003 \\
60--70 & 0.0197  &  0.0083  &  0.0013  &  0.0013   \\
70--80 & 0.0081& 0.0029 & 0.0008 & 0.0006 \\
80--90 & 0.0048 & 0.0021 &  0.0006 &  0.0004  \\
90--100 & 0.0024 & 0.0009 &  0.0005 & 0.0002 \\
\hline 
\end{tabular}
\end{center}
\label{tab:mucrosssection}
\end{table}

The measured heavy-flavour cross-sections 
are compared to 
the FONLL calculations, with a
rigorous evaluation of the associated uncertainty shown as a band in Fig.~\ref{fig:FullComparison}.
The theoretical uncertainties originate from several different sources. 
The dominant contribution comes from the renormalisation and factorisation scales 
(up to 35\% at low \pt).\footnote{The renormalisation ($\mu_{\rm R}$)
and factorisation ($\mu_{\rm F}$)  scales are defined as $\mu_{\rm R,F} = \xi_{\rm {R,F}}\sqrt{p_\mathrm{T}^2+m_Q^2}$. The central value is computed using 
$\xi_{\rm {R,F}} = 1$ while the scale uncertainty is determined by
changing  the scales independently within $0.5 < \xi_{\rm {R,F}} < 2.0$ while
keeping the ratio $0.5 < \xi_{\rm {R}}/\xi_{\rm {F}} < 2.0$.}
The uncertainty on the heavy quark masses contributes up to 9\% at
low \pt{\footnote{The heavy quark masses are set to $m_b = 4.75 \pm 0.25$ GeV and $m_c = 1.5 \pm
0.2$~GeV.}}, and the PDF-related uncertainty 
(taken from the  {\tt CTEQ6.6} error set)
is below 8\% over the whole~\pt~ range.
Uncertainties arising from the value of $\alpha_s$ and
on the non-perturbative fragmentation function are found to be small, 
approximately 1\% and less than 5\%~\cite{FONLL-HQ0} respectively.
The total uncertainty, dominated by the renormalisation and factorisation scales, 
is in the approximate range 20-40\%, decreasing with \pt.
The electron and muon results are seen to be fully compatible with the overall FONLL uncertainty bands.

The results are also compared to the NLO predictions of the {\tt POWHEG} program, 
interfaced to either {\tt PYTHIA} or {\tt HERWIG} for the parton shower simulation, 
and to the LO plus parton shower predictions of {\tt PYTHIA}.
Whereas {\tt POWHEG+PYTHIA} agrees well with the FONLL predictions,  
{\tt POWHEG+HERWIG} predicts a significantly lower total cross-section. 
Less than half of this difference may be accounted for by the
different heavy-flavour hadron decay models, checked by
implementing  a common decay simulation, {\tt EVTGEN} \cite{EVTGEN},  for
both showering and hadronisation programs. 
 {\tt PYTHIA} (LO) describes the \pt-dependence well but predicts approximately a factor two higher total 
cross-section.

Comparisons are also made to the NLO central value expectation obtained from the FONLL
program by excluding the NLL resummation part of the pQCD calculation.
As shown in Fig.~\ref{fig:FullComparison} (right), the data deviate significantly from the NLO prediction, showing 
sensitivity to the NLL resummation term in the pQCD calculation for the first time in heavy-flavor production at hadron colliders.

\clearpage

\section{Conclusions}

The differential cross-sections of electrons and muons arising from heavy-flavour  production 
have been measured and  
found to be in good agreement in the transverse momentum range 7 $<$ \pt~$ <$ 26 GeV and 
pseudorapidity region  $|\eta|<2.0$ (excluding $1.37<|\eta|<1.52$). 
The inclusive differential cross-section of muon production has also been measured in the 
extended \pt~range 4 $< $ \pt~$ <$ 100 GeV within  $|\eta|<2.5$.

The theoretical predictions for heavy-flavour production from the FONLL computation
are in  good agreement with the electron and muon measurements. 
Good agreement is also seen with the predictions of 
{\tt POWHEG+PYTHIA}, although {\tt POWHEG+HERWIG} predicts a significantly lower total cross-section. 
 {\tt PYTHIA} describes the \pt-dependence well but predicts approximately a factor two higher total 
cross-section.
For muons with \pt~$> 25$ GeV  a deviation from the NLO central prediction is seen, indicating sensitivity of the heavy-flavour production data to the NLL high-\pt~resummation terms.

\section{Acknowledgments}
We thank Matteo Cacciari for supplying the theoretical predictions from FONLL, within our acceptance cuts, 
and for many useful discussions.

We also thank CERN for the very successful operation of the LHC, as well as the support staff from our institutions 
without whom ATLAS could not be operated efficiently. We acknowledge the support of ANPCyT, Argentina; YerPhI, Armenia; 
ARC, Australia; BMWF, Austria; ANAS, Azerbaijan; SSTC, Belarus; CNPq and FAPESP, Brazil; NSERC, NRC and CFI, Canada; 
CERN; CONICYT, Chile; CAS, MOST and NSFC, China; COL3 CIENCIAS, Colombia; MSMT CR, MPO CR and VSC CR, Czech Republic; 
DNRF, DNSRC and Lundbeck Foundation, Denmark; ARTEMIS, European Union; IN2P3-CNRS, CEA-DSM/IRFU, France; GNAS, Georgia; 
BMBF, DFG, HGF, MPG and AvH Foundation, Germany; GSRT, Greece; ISF, MINERVA, GIF, DIP and Benoziyo Center, Israel; 
INFN, Italy; MEXT and JSPS, Japan; CNRST, Morocco; FOM and NWO, Netherlands; RCN, Norway; MNiSW, Poland; 
GRICES and FCT, Portugal; MERYS (MECTS), Romania; MES of Russia and ROSATOM, Russian Federation; JINR; MSTD, Serbia; 
MSSR, Slovakia; ARRS and MVZT, Slovenia; DST/NRF, South Africa; MICINN, Spain; SRC and Wallenberg Foundation, Sweden; 
SER, SNSF and Cantons of Bern and Geneva, Switzerland; NSC, Taiwan; TAEK, Turkey; STFC, the Royal Society and Leverhulme 
Trust, United Kingdom; DOE and NSF, United States of America. 

The crucial computing support from all WLCG partners is acknowledged gratefully, in particular from CERN and
 the ATLAS Tier-1 facilities at TRIUMF (Canada), NDGF (Denmark, Norway, Sweden), CC3IN2P3 (France), KIT/GridKA 
(Germany), INFN-CNAF (Italy), NL-T1 (Netherlands), PIC (Spain), ASGC (Taiwan), RAL (UK) and BNL (USA) and in the 
Tier-2 facilities worldwide.

\clearpage

\bibliographystyle{model1-num-names}
\bibliography{Paper}

\clearpage 
\begin{flushleft}
{\Large The ATLAS Collaboration}

\bigskip

G.~Aad$^{\rm 48}$,
B.~Abbott$^{\rm 111}$,
J.~Abdallah$^{\rm 11}$,
A.A.~Abdelalim$^{\rm 49}$,
A.~Abdesselam$^{\rm 118}$,
O.~Abdinov$^{\rm 10}$,
B.~Abi$^{\rm 112}$,
M.~Abolins$^{\rm 88}$,
H.~Abramowicz$^{\rm 153}$,
H.~Abreu$^{\rm 115}$
% ...
\bigskip

% ...
$^{10}$ Institute of Physics, Azerbaijan Academy of Sciences, Baku, Azerbaijan\\
$^{11}$ Institut de F\'isica d'Altes Energies and Universitat Aut\`onoma  de Barcelona and ICREA, Barcelona, Spain\\
% ...
$^{48}$ Fakult\"{a}t f\"{u}r Mathematik und Physik, Albert-Ludwigs-Universit\"{a}t, Freiburg i.Br., Germany\\
$^{49}$ Section de Physique, Universit\'e de Gen\`eve, Geneva, Switzerland\\
% ...
$^{88}$ Department of Physics and Astronomy, Michigan State University, East Lansing MI, United States of America\\
% ...
$^{111}$ Homer L. Dodge Department of Physics and Astronomy, University of Oklahoma, Norman OK, United States of America\\
$^{112}$ Department of Physics, Oklahoma State University, Stillwater OK, United States of America\\
% ...
$^{115}$ LAL, Univ. Paris-Sud and CNRS/IN2P3, Orsay, France\\
% ...
$^{118}$ Department of Physics, Oxford University, Oxford, United Kingdom\\
% ...
$^{153}$ Raymond and Beverly Sackler School of Physics and Astronomy, Tel Aviv University, Tel Aviv, Israel\\
% ...
\end{flushleft}


\end{document}